\documentstyle[aps,prl,psfig,amssymb]{revtex}
\begin{document}
\draft
\twocolumn[\hsize\textwidth\columnwidth\hsize\csname
@twocolumnfalse\endcsname
\title{
Emergence of a small world from local interactions: 
\\
Modeling acquaintance networks
}
\author{J\"orn Davidsen, Holger Ebel and Stefan Bornholdt\cite{email}}
\address{
Institut f\"ur Theoretische Physik,
Universit\"at Kiel, Leibnizstr.\ 15, D-24098 Kiel, Germany}
\date{\today}
\maketitle
\begin{abstract}
How does one make acquaintances? A simple observation from everyday 
experience is that often one of our acquaintances introduces us to 
one of his acquaintances. Such a simple triangle interaction may be 
viewed as the basis of the evolution of many social networks. Here, 
it is demonstrated that this assumption is sufficient to reproduce 
major non-trivial features of social networks: Short path length, 
high clustering, and scale-free or exponential link distributions.  
\end{abstract}
]

A remarkable feature of many complex systems is the occurrence of 
large and stable network structures, as, for example, networks on 
the protein or gene level, ecological webs, communication networks, 
and social networks \cite{strogatz01,albert01,dorogovtsev01}. 
Already simple models based on disordered networks are quite successful 
in describing basic properties of such systems. When addressing 
topological properties, however, neither random networks nor regular 
lattices provide an adequate framework to model characteristic features. 
A helpful concept along this line is the idea of ``small-world networks'' 
introduced by Watts and Strogatz \cite{watts98,watts99}, 
which initiated an avalanche of scientific activity in this field 
\cite{newman00a,barthelemy99,newman99,monasson99,barrat00,newman00b}. 
Small-world networks interpolate between the two limiting cases of 
regular lattices with high local clustering and random graphs with 
short distances between nodes. High clustering means that, if node 
$A$ is linked to node $B$, and $B$ is linked to node $C$, there is 
an increased probability that $A$ will also be linked to $C$. 
Another useful measure is the distance between two nodes, defined 
as the number of edges along the shortest path connecting them.
A network is called a ``small-world network'' if it exhibits the 
following two characteristic properties \cite{watts98,amaral00}: 
(i) high clustering; and (ii) a small average shortest path 
between two random nodes (the diameter of the network), scaling 
logarithmically with the number of nodes. Thus, any two nodes 
in the network are connected through only a small number of links. 
The most popular manifestation of a small world is known as 
``six degrees of separation'', a postulate by the social 
psychologist Stanley Milgram \cite{milgram67} stating that most 
pairs of people in the United States can be connected through a 
path of only about six acquaintances \cite{kochen89}.

Let us here focus on social networks, and acquaintance networks 
in particular, which are typical examples for the small-world 
property \cite{strogatz01,albert01,amaral00}. First of all, what 
does the concept of ``small-world networks'' tell us about real 
world systems? In its original definition \cite{watts98,watts99} 
it served as an elegant toy model, demonstrating the consequences 
of high clustering and short path length. However, as these networks 
are derived from regular graphs, their applicability to real world 
systems is very limited. In particular, how a network in a natural 
system forms a small-world topology dynamically, often starting 
from a completely random structure, remains unexplained. The main 
goal of this paper is to provide one possible answer to this problem. 

A similar problem of dynamical origin is faced (and much progress 
has been made) in a different, but not completely unrelated field: 
The dynamics of scale-free networks. Scale-free properties are 
commonly studied in diverse contexts from, e.g., the stability 
of the internet \cite{albert00b} to the spreading of epidemics 
\cite{pastor01}, and are observed in some social networks   
\cite{strogatz01,albert01,dorogovtsev01,newman01b,barabasi01b}. 
The origin of scale-free properties is well understood in terms
of interactions that generate this topology dynamically, 
e.g., on the basis of network growth and preferential linking
\cite{albert01,dorogovtsev01,bornholdt01}. While these models 
generate scale-free structures, they do not, in general, 
lead to clustering and are therefore of limited use 
when modeling social networks. 

In this paper, an attempt is made to unify ideas from the two 
worlds of ``small-world networks'' and ``scale-free networks'' 
which may help understanding social networks, and how a 
small-world structure can emerge dynamically. In particular, 
a simple dynamical model for the evolution of acquaintance networks
is studied. 
It generates highly clustered networks with small average path 
lengths which scale logarithmically with network size. 
Furthermore, for small death-and-birth rates of nodes this 
model converges towards scale-free degree distributions, 
in addition to its small-world behavior. Basic ingredients are 
a local connection rule based on ``transitive linking'', and 
a finite age of nodes.

To be specific, let us formulate a model of an acquaintance network 
with a fixed number $N$ of nodes (as persons) and undirected links 
between those pairs of nodes that represent people who know each other. 
Acquaintance networks evolve, with new acquaintances forming between 
individuals, and old ones dying. Let us assume that people are usually 
introduced to each other by a common acquaintance and that the network 
is formed only by people who are still alive. The dynamics is defined 
as follows: 
\medskip

(i) One randomly chosen person picks two random acquaintances of his, 
and introduces them to each another. If they have not met before, a new 
link between them is formed. In case the person chosen has less than 
two acquaintances, he introduces himself to one other random person. 

(ii) With probability $p$, one randomly chosen person is removed from 
the network, including all links connected to this node, and replaced 
by a new person with one randomly chosen acquaintance.
\medskip
 
\noindent 
These steps are then iterated. Note that the number of nodes remains 
constant, neglecting fluctuations in the number of individuals in 
acquaintance networks. The finite age implies that the network reaches 
a stationary state which is an approximation of the behavior of many social 
networks, and is in contrast to most models based on network growth
\cite{albert01,barabasi01b,klemm01}. The probability $p$ 
determines the separation 
of the two timescales in the model.
In general, the rate at which people make social contacts can be as 
short as minutes or hours, while the timescale on which people join 
or leave the network may be as long as years or decades. In the 
following, 
we therefore focus on the regime $p \ll 1$.

\begin{figure}[hbt]
\centerline{\psfig{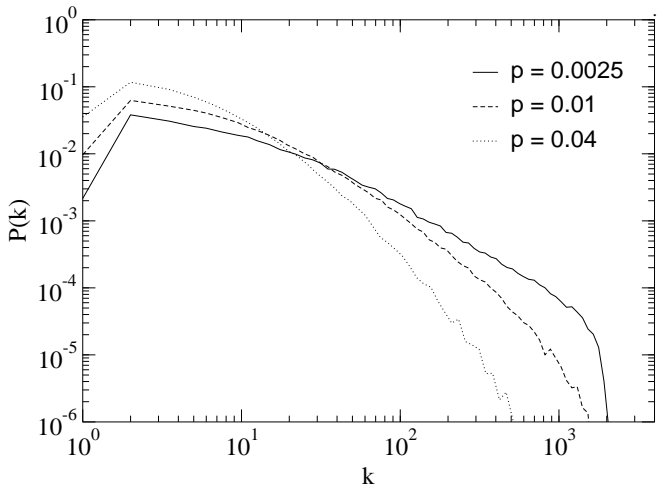}}
\vspace{0.25cm}
\caption{
The degree distribution $P(k)$ of the transitive
linking model in the statistically stationary state.
The distribution exhibits a power-law regime
for small $p$, with an exponent of 1.35 for $p=0.0025$.
Note that the distribution is largely insensitive to
system size $N$, which here is $N = 7000$. The
exponential cutoff is a result of the finite age of nodes.}
\label{degree_dist}
\end{figure}

Once the network reaches a statistically stationary state, one of its 
characteristic quantities is the degree distribution $P(k)$ of the 
network. In Fig. \ref{degree_dist}, the degree distribution is 
shown for different values of $p$. 
Due to the limited lifetime 
of persons in the network, the observed numbers of acquaintances 
of different persons do not grow forever, but rather fall into some 
finite range. This can be seen in the degree distribution where
the cut-off at high $k$ results from the finite age of nodes. 
In the regime $p \ll 1$, the degree distribution is dominated by 
transitive linking (i), resulting in a power-law range which 
increases in size with decreasing $p$. For larger values of $p$, 
the Poissonian death process (ii) competes with the transitive 
linking process (i), resulting in a stretched exponential range 
in the degree distribution until, in the case $p\approx 1$, 
the random linking of (ii) dominates with its Poissonian dynamics. 
Therefore, the above model generates degree distributions 
spanning scale-free and exponential regimes that are also observed
in the statistics of social networks. For large enough network 
size $N$, the specific distribution $P(k)$ only depends on the 
single free parameter of the model $p$. From experimental data
one can estimate $p$ as well, which is typically very small 
$p\ll 1$ such that the two timescales of the network dynamics
are well separated.  

As already noted above, ``small-world networks'' are 
characterized by a high degree of clustering $C$ and a 
small average path length $\ell$ which scales logarithmically 
with the number of nodes. The degree of clustering is measured 
by the clustering coefficient defined as follows: 
For a distinct node $i$, the clustering coefficient $C_i$ 
is given by the ratio of existing links $E_i$ between its 
$k_i$ neighbors to the possible number of such connections 
$\frac{1}{2} k_i (k_i-1)$. Then the clustering coefficient 
$C$ of the network is defined as the average over all nodes
\begin{equation}
C = \langle C_i \rangle_i = \left \langle \frac{2 E_i}{k_i(k_i-1)} 
\right \rangle_i.
\end{equation}
In Table \ref{table_clust}, the clustering coefficient of the
above model is shown for different values of $p$. 
\begin{table}[htb] 
\caption{Clustering coefficient for different values of $p$
and a network size of $N = 7000$. $C'$ is an upper bound
for the average clustering coefficient of a network with
the same degree distribution without transitive linking.
$C_{\em rand}$ is the clustering coefficient of a random
network of the same size and with the same average degree
$\langle k \rangle$.}
\label{table_clust}
\begin{tabular}{|d|d|d|d|d|d|}
$p$ & $\langle k \rangle$ & $\langle k^2
\rangle$ & $C$ & $C'$ & $C_{\em rand}$\\
\hline
0.04 &  14.9 &  912 &  0.45 &  0.036 &  0.0021\\
0.01 &  49.1 &  13744 &  0.52 &  0.29 &  0.0070\\
0.0025 &  149.2 &  99436 & 0.63 & 0.43 & 0.021\\
\end{tabular}
\end{table}
In comparison 
to the clustering of a random network with same size and same 
mean degree $C_{\em rand}$, the model coefficient $C$ is 
consistently of much larger size. The clustering coefficient 
of a random network, i.e. a network with constant probability 
of linking each pair of nodes $p_{\em link}=\langle k \rangle 
/ (N-1)$ and, therefore, a Poisson distribution of the node 
degree, is just this probability $C_{\em rand}=p_{\em link}$. 
Obviously, $C_{\em rand}$ is proportional to the mean degree 
$\langle k \rangle$ for constant network size. 
For further comparison, let us derive an estimate $C'$ for 
an upper bound of the average 
clustering coefficient of a network with the same degree 
distribution, but randomly assigned links. Thus, $C'$ provides
an upper bound on the average clustering which one would expect solely 
from the degree distribution while neglecting transitive linking 
of the model. Using the generating function approach for graphs 
with arbitrary degree distributions \cite{newman00} and assuming 
that fluctuations of the average degree of the neighborhood of 
a node can be neglected, an upper bound $C'$ can be derived in
terms of the first two moments:
\begin{equation}
C' = \frac{1}{\langle k \rangle N} 
\left (\frac{\langle k^2 \rangle}{\langle k \rangle}-1\right 
)^2.\label{upper_bound}
\end{equation}
This result holds exactly in the case of the Poissonian degree
distribution of a random network ($C_{\em rand} = C'$). 
As Table \ref{table_clust} shows, the network generated by the 
model exhibits an even higher average clustering coefficient than a 
network with links distributed randomly according to the same degree 
distribution. In particular, the network is much more clustered 
than a random network with a Poisson distribution as required 
for the small-world property. Furthermore, the clustering is not as 
much dependent on the mean degree $\langle k \rangle$ as $C'$ of 
a network with the same degree distribution but no transitive linking.

\begin{figure}[hbt]
\centerline{\psfig{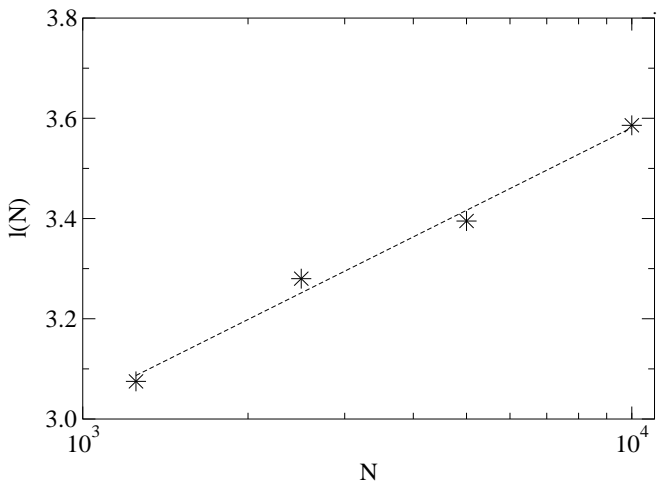}}
\vspace{0.25cm}
\caption{The average path length $\ell(N)$ as a function of
system size $N$ on a semilogarithmic scale ($p = 0.04$).
The data are in good agreement with a logarithmic
fit (straight line). }
\label{path_length}
\end{figure}

The scaling of the average path length with system size is shown 
in Fig. \ref{path_length}. 
The data are consistent with a logarithmic 
behavior and, thus, our model meets the second requirement for a 
small-world behavior, as well. Also, this is what one expects in
the framework of a random graph with arbitrary degree distribution 
\cite{newman00}. Moreover, we can compare the average path length 
of our model with the path length of a network with the same degree 
distribution and randomly assigned links. The generating function 
approach can again be used to estimate the average path length $\ell$ 
of such a network. For a broad distribution this leads to:
\begin{equation}
\ell' \approx \frac{\log \left (\frac{N}{\langle k \rangle}
\right )}{\log \left (\frac{\langle
 k^2 \rangle - \langle k \rangle}{\langle k \rangle}\right )}+1.
\end{equation}
For the Poisson distribution of a random network, one obtains:
\begin{equation}
\ell_{\em rand} \approx  \frac{\log N}{\log \langle k \rangle}.
\end{equation}
Note that only $N$, $\langle k \rangle$ and $\langle k^2 \rangle$ are 
used to estimate $\ell$ as also used in the derivation of 
(\ref{upper_bound}).
With the help of Table \ref{table_clust} one finds that $\ell' 
\approx 1.59$ and $\ell_{\em rand} \approx 1.77$ for $p = 0.0025$. 
Numerical simulations of our model similarly yield a very short path 
length of $\ell = 2.38$ which is further evidence that the simple 
linking rule of our model leads to small-world behavior. Intuitively, 
$\ell' < \ell_{\em rand}$ follows from the highly connected ``hubs'' 
present in the scale-free networks for $p=0.0025$. The fact that $\ell$ 
is slightly larger than $\ell_{\em rand}$ is due to the fact that 
in step (i) of the model, many links are used to build highly 
clustered neighborhoods. This price to pay for clustering, however, 
only slightly affects the small overall mean path length.    

One example for an observed small-world effect is the network of 
coauthorships between physicists in high energy physics 
\cite{newman01b}. Nodes are researchers who are connected if they 
have co-authored a paper. In a recent study of the publications in 
the SPIRES database over the five year period 1995-1999, a graph 
was reconstructed from the data and analyzed \cite{newman01b,newman01c}. 
The resulting network consists of $55,627$ nodes with a mean degree 
$\langle k_S \rangle = 173$, a mean shortest path length $\ell_S=4.0$, 
and a very high clustering coefficient $C_S=0.726$. The degree 
distribution is consistent with a power-law of exponent $-1.2$ 
\cite{newman01b}. From these data one can derive $C'$ and $\ell'$ 
for a network with the same degree distribution but random links 
to $C'=0.19$ and $\ell'=1.81$. These numbers show that the real network 
exhibits clustering which is very much higher than would be expected 
from the degree distribution alone. Also, the path length is short 
but still larger than for a randomly linked network of the same 
degree distribution. Using the logarithmic scaling for $\ell$, 
the data of the example agree with the values of the above model. 
The basic assumptions made in the model are met by the dataset, as 
the number of researchers in the sample is to a good approximation 
stationary, and as the small rate of researchers entering or leaving 
the system during the time frame of the sample justifies 
the regime of small $p\ll 1$. 

The previous example demonstrates, how our model  
can be applied to a social network in the dynamically 
stationary state. Moreover, the model studied here is also able to 
accommodate other small-world scenarios, e.g., without a scale-free 
degree distribution \cite{amaral00} as the particular shape of 
the distribution varies, depending on the turnover rate $p$.  
The question of the origin of small-world behavior in social systems 
has led to many different approaches as well. In an interesting 
model Mathias and Gopal \cite{mathias01} showed that a small 
world topology can arise from the combined optimization of network 
distance and physical distance. Applications of this principle, 
however, are more likely to be found in the field of transportation 
networks rather than acquaintance networks considered here. 
A more similar approach to the study presented here has been taken
by Jin et al. \cite{jin01}, who study a model which shares a mechanism 
similar to transitive linking as defined here. Otherwise, however, 
it is more complicated than we feel it needs to be, at least for 
some classes of social networks. Also its upper limit on the degree 
of a node, motivated by one specific trait of some acquaintance 
networks, otherwise makes the model less suitable to meet 
experimental data of social networks which exhibit broad 
degree distributions.  

In conclusion, a simple dynamical model for the emergence of network 
structures in social systems has been studied. It is based on a local 
linking rule which connects nodes who share a common neighbor, 
as well as on a slow turnover of nodes and links in the system. 
The network approaches a dynamically stationary state, with high 
clustering and small average path lengths which scale 
logarithmically with system size. Depending on a single free 
parameter, the turnover rate of nodes, this model interpolates 
between networks with a scale-free and an exponential degree 
distribution, both of which are observed in experimental data of 
social networks.

\end{document}